\documentclass[reqno, 12pt]{article}
\usepackage{jheppub,graphicx}

\DeclareMathOperator{\e}{\epsilon}
\DeclareMathOperator{\p}{\partial}
\DeclareMathOperator{\Tr}{Tr}

\newcommand{\be}{\begin{equation}}
\newcommand{\ee}{\end{equation}}

\begin{document}
\subheader{SU-ITP-11/53}
\title{Holographic Entanglement Entropy and Fermi Surfaces}
\author[\dagger]{Edgar Shaghoulian}
\note[$^{\dagger}$]{edgars@stanford.edu}
\affiliation{Stanford Institute for Theoretical Physics, Department of Physics, Stanford University, Stanford, CA 94305}
\abstract{The entanglement entropy in theories with a Fermi surface is known to produce a logarithmic violation of the usual area law behavior. We explore the possibility of producing this logarithmic violation holographically by analyzing the IR regions of the bulk geometries dual to such theories. The geometry of Ogawa, Takayanagi, and Ugajin is explored and shown to have a null curvature singularity for all values of parameters, except for dynamical critical exponent 3/2 in four dimensions. The results are extended to general hyperscaling violation exponent. We explore strings propagating through the singularity and show that they become infinitely excited, suggesting the singularity is not resolved by stringy effects and \emph{may} become a full-fledged ``stringularity." An Einstein-Maxwell-dilaton embedding of the nonsingular geometry is exhibited where the dilaton asymptotes to a constant in the IR. The unique nonsingular geometry in any given number of dimensions is proposed as a model to study the $T=0$ limit of a theory with a Fermi surface.}
\maketitle

\begin{section}
{Introduction}\label{intro}
\end{section}
Holography is a powerful tool that has recently been employed to study quantum field theories that admit a Fermi surface. The renormalization group flow of the quantum field theory is geometrized in these models as a flow from large distances to short distances in the bulk.

Another tool to study quantum field theories, which has recently been given a holographic interpretation that we will explore in this paper, is the geometric entropy. The geometric entropy is defined as the entanglement entropy between two regions $A$ and $B$, where $A\cup B$ is the configuration space of the entire system. The Hilbert space then admits a decomposition along this separation, $\mathcal{H}=\mathcal{H}_A\bigotimes\mathcal{H}_B$. The entanglement entropy is computed as 
\be
\label{crap} S_A=-\Tr_A \rho_A \log \rho_A,
\end{equation}
where 
\begin{equation}
\rho_A = \Tr_B \rho
\end{equation}
with $\rho$ the density matrix of the whole system. In general the geometric entropy is a time-dependent quantity, but in this paper we will focus solely on time-independent cases. It was shown that for an arbitrary spatial division of a local quantum field theory, the geometric entropy (henceforth referred to as the entanglement entropy) between $A$ and $B$ for the ground state density matrix scales as the area of the dividing region \cite{Bombelli:1986rw, Srednicki:1993im}:
\be
S_A \sim \frac{\textrm{Area}(\partial A)}{\e^{d-1}}+\textrm{subleading terms}\ee
for a $d+1$ dimensional quantum field theory. $\e$ is the UV cutoff of the theory and $\partial A$ is the dividing manifold with Area$(\partial A)\sim L^{d-1}$ for some characteristic length scale $L$. This behavior degenerates in the case $d=1$ to a logarithmic scaling: $S_A \sim \log (L/\e)$ where $L$ is the length of the region $A$. 

This area law is to be contrasted with the extensive behavior of thermodynamic entropy and invites speculation on the connection to black holes. It is intuitively understood by observing that for local theories in their ground state, the interactions occur most strongly across the boundary. There is a violation of this area law when the quantum field theory admits a Fermi surface; in that case, the entanglement entropy scales as \cite{Wolf:2006zzb, Gioev:2006zz}
\be
S_A \sim \left(\frac{L}{\e}\right)^{d-1}\log\frac{L}{\e}+\textrm{subleading terms},
\ee
where $L$ is the characteristic length scale of $\partial A$. This logarithmic behavior is understood as coming from the 2D CFT that approximates the excitations normal to a Fermi surface.

Entanglement entropy has recently been given a holographic interpretation in the AdS$_{d+2}$/CFT$_{d+1}$ setup \cite{Ryu:2006bv}. We imagine a time-independent bulk geometry that is dual to some quantum field theory on the boundary. We begin by taking a time slice of the boundary field theory. In the time-independent case this corresponds to a time slice of the bulk; we work completely within this slice. We split the spatial manifold on the boundary of this slice into two pieces $A$ and $B$ with a hypersurface $\partial A$. To compute the entanglement entropy between these two regions, we first find the minimal surface on this time slice in the bulk that is anchored at the boundary on $\partial A$. We will follow the convention and call this minimal surface $\gamma_A$. Then the entanglement entropy is given by 
\be
S_A = \frac{\textrm{Area}(\gamma_A)}{4G_N},
\ee
where $G_N$ refers to the bulk Newton's constant. This proposal trivially reproduces the ``area law," which we show explicitly in Appendix \ref{holo} (for a review, see \cite{Nishioka:2009un}). 

In this paper, we wish to study the cases in which the holographic entanglement entropy formula can reproduce the logarithmic scaling characteristic of the existence of a Fermi surface. We will analyze the geometry of \cite{Ogawa:2011bz} (see also \cite{Huijse:2011ef}), which we call the OTU geometry after the authors' surnames, and other cases below. We will show that the geometry generically has a null curvature singularity, except for a specific set of parameters. The geometry can be modeled by a singular plane wave, and we will show that strings in this background become infinitely excited at the singularity, suggesting that the geometry is singular in the sense of string theory as well. We will find that geometries that admit an IR region with a time slice equal to EAdS$_2\times \mathbb{R}^{d-1}$ are the simplest geometries that can reproduce the correct scaling, although their boundary interpretation is unclear. Geometries such as the electron star and the extremal Reissner-Nordstrom black hole will not produce the violation.

Those who wish to bypass the operatic crescendo of this paper may go directly to Sections 3 and 4, which concern the singularities and gravitational embedding of the OTU geometry and form the primary results. Some preliminary applications of the holographic entanglement entropy formula can be found in Appendix \ref{holo}.\\

\begin{section}
{Area Law Violation}
\end{section}

In this section we analyze what criteria are sufficient to violate the area-law scaling and produce the logarithmic correction characteristic of Fermi surfaces. We first note a seeming problem, which is that $\emph{any}$ asymptotically AdS$_{d+2}$ geometry will $\emph{always}$ produce the area-law scaling $L^{d-1}/\e^{d-1}$, since this contribution comes from the piece of the minimal surface out near the boundary of the spacetime. However, it seems clear that the violation should occur in terms of an $\emph{effective}$ length and an $\emph{effective}$ cutoff (which we call $L_{\textrm{eff}}$ and $\e_{\textrm{eff}}$) coming from the piece of the minimal surface probing deep into the bulk.\footnote{As in \cite{Ogawa:2011bz}, we expect $\epsilon_{\textrm{eff}}\sim k_F^{-1}$ (this parameter will sometimes be set to 1) for the Fermi momentum $k_F$, so statements concerning the IR geometry are meant to imply the limit $z\gg \epsilon_{\textrm{eff}}$.} This is because in holographic descriptions of Fermi surfaces, we expect to begin with a CFT$_{d+1}$ in the UV of our boundary theory and flow to a theory with a Fermi surface in the IR. The IR of the boundary theory corresponds to regions deep in the bulk geometry, so the logarithmic violation should be produced there. Of course, there are subtleties in this interpretation, but we will adopt it as our operating assumption to see what answers we get.

Instead of working in the full geometry and considering large minimal surfaces that probe deep into the geometry, we can analyze the IR region of the geometry alternatively by pushing the cutoff deep into the bulk and excising the UV region. Then, the geometry on our boundary theory slice will be different and we can hope to get a different type of scaling as we approach this new ``UV" boundary; this will be made precise in an example below. We will find that a geometry that becomes AdS$_3 \times \mathbb{R}^{d-1}$ deep in the bulk (or, more generally, its time slice becomes EAdS$_2\times\mathbb{R}^{d-1}$) will satisfy the logarithmic violation with respect to $L_{\textrm{eff}}$ and $\e_{\textrm{eff}}$.\footnote{$\e_{\textrm{eff}}$ refers to the new location of the boundary slice, while $L_{\textrm{eff}}$ refers to the characteristic size of $\partial A$ on this slice. See Appendix \ref{fullgeometry} for an example of how we expect $\e_{\textrm{eff}}$ and $L_{\textrm{eff}}$ to arise from the full geometry.} This is simply because the AdS$_3$ piece will give the logarithm we are looking for, while the $\mathbb{R}^{d-1}$ piece will give the ``extensive" $L_{\textrm{eff}}^{d-1}/\e_{\textrm{eff}}^{d-1}$ piece. They will combine to give $\left(L_{\textrm{eff}}^{d-1}/\e_{\textrm{eff}}^{d-1}\right)\log \left(L_{\textrm{eff}}/\e_{\textrm{eff}}\right)$, which is the result we hope for in an asymptotically AdS$_{d+2}$ geometry. We will also see that the electron star and extremal Reissner-Nordstrom black hole geometries do not give the correct scaling. Adopting a certain metric ansatz, the authors of \cite{Ogawa:2011bz} produce general geometries which contain the logarithmic violation; we will present the geometry and analyze its singularity nature in Section \ref{singular}.

Viewed in this way, working in the IR geometry is simply a shorthand to extract the IR pieces of the full minimal surface. This approach misses important UV-IR mixing terms since the UV part of the geometry is excised. It is possible that logarithmic violations can come from these intermediate regions.

When working in the IR geometries, we will consider them geometries in their own right and apply the holographic entanglement entropy prescription from their ``boundaries." When we use this language it is simply shorthand for the procedure involving pushing the boundary slice deep into the IR region. This distinction is important because the IR boundaries we consider are not meant to be considered as holographic screens (or as conformal boundaries), and as stated above serve only as a tool to extract the IR contribution to the entanglement entropy.\\

\begin{subsection}
{AdS$_3\times\mathbb{R}^{d-1}$}\label{mystuff}
\end{subsection}

In this section we will analyze\footnote{The entangling properties of this spacetime were also studied in \cite{Swingle:2010bt}} AdS$_3\times\mathbb{R}^{d-1}$ (again, any geometry with a time slice of EAdS$_2\times\mathbb{R}^{d-1}$ will do the trick, but we stick to this geometry for concreteness), which we hope to appear in the IR of some asymptotically AdS$_{d+2}$ geometry. We write this IR metric as

\be
ds_{IR}^2=\frac{1}{z^2}(-dt^2+dz^2+dy^2)+\frac{1}{\e_{\textrm{eff}}^2}dx_i^2
\ee
for $i=1,2,\dots d-1$ and $R_{\textrm{AdS}}=1$. We follow the warping to the regulated ``boundary" at $z=\e_{\textrm{eff}}$. Since the IR metric is a simple product manifold, the minimal surfaces will be product manifolds as well. We already know the AdS$_3$ answer for a minimal surface (in this case just a spacelike geodesic); for a length $L_{\textrm{eff}}$ on the boundary it is given by 
\be
(z,y)=\frac{L_{\textrm{eff}}}{2}(\sin s, \cos s),
\ee
which has length 
\be
\textrm{Length}\sim \log \frac{L_{\textrm{eff}}}{\e_{\textrm{eff}}}.
\ee
For the full boundary geometry including the $\mathbb{R}^{d-1}$ piece, we consider a $d-1$ dimensional box of side length $L_{\textrm{eff}}$ separating a time slice of the boundary into regions $A$ in the interior of the box and $B$ representing the exterior. Restricting to the $\mathbb{R}^{d-1}$ part of the geometry, this becomes a $d-2$ dimensional box of side length $L_{\textrm{eff}}$. The minimal surface anchored at the boundary on this $d-2$ dimensional box is exceedingly simple to find; you simply fill in the interior of the box! Since no warping appears, the surface remains on the boundary slice and does not reach into the bulk. The area of this minimal surface scales as 
\be
\textrm{Area} \sim \frac{L_{\textrm{eff}}^{d-1}}{\e_{\textrm{eff}}^{d-1}},
\ee
which is simply the extensive scaling observed in Appendix \ref{flat}.
Considering the product of these two minimal surfaces, we get the appropriate scaling 
\be
\textrm{Area} \sim \frac{L_{\textrm{eff}}^{d-1}}{\e_{\textrm{eff}}^{d-1}}\log\frac{L_{\textrm{eff}}}{\e_{\textrm{eff}}}.
\ee
Notice that all we required in this analysis is that a time slice of the geometry be EAdS$_2\times\mathbb{R}^{d-1}$. For example, modifying to Lifshitz scaling will not change these results, as long as it is a Lif$_3$ piece analogous to our AdS$_3$ piece, which will keep the same induced geometry on a time slice.


The physical interpretation of the boundary theory is less clear. The AdS$_3$ factor suggests that the specific heat (and entropy density) depend linearly on temperature, just as a conventional Fermi liquid. A troubling aspect of this geometry is that it is usually produced by turning on magnetic fields, in which case the low energy theory becomes effectively 1+1-dimensional (describing the physics of the Lowest Landau Level), and the logarithm we are picking up is simply that of the 1+1-dimensional theory. Such a model is given by the magnetic brane solution of \cite{D'Hoker:2009mm, Almuhairi:2011ws}. At this level, this geometry is presented simply as a proof of principle for the existence of the logarithmic violation, before moving on to the more general geometries of \cite{Ogawa:2011bz}.\\

\begin{subsection}
{Other Fermi Surface Geometries}\label{other}
\end{subsection}
In this section we analyze some of the other geometries that are purported to have Fermi surface duals. One model is the extremal Reissner-Nordstrom black hole \cite{Faulkner:2009wj, Liu:2009dm, Lee:2008xf, Cubrovic:2009ye}
\be
ds^2 = \frac{R^2}{z^2}\left(-f(z)dt^2+\frac{dz^2}{f(z)}+dx_i^2\right),
\ee
where 
\be
f(z)=1-4\left(\frac{z}{z_0}\right)^3+3\left(\frac{z}{z_0}\right)^4
\ee
is the emblackening factor. Expanding this geometry in the near horizon limit (around $z=z_0$) and rescaling coordinates one gets 
\be
ds^2=\frac{R^2}{p^2}(-dt^2+dp^2)+dx_i^2,
\ee
which is simply AdS$_2 \times \mathbb{R}^d$. The induced metric on a constant time slice can be written as 
\be
ds^2 = \frac{R^2}{p^2}dp^2+dx_i^2=d\bar{p}^2+dx_i^2,
\ee
which is simply $\mathbb{R}^{d+1}$. This will produce an extensive entanglement entropy contribution $L_{\textrm{eff}}^d/\e_{\textrm{eff}}^d$ as argued in Appendix \ref{flat}. The fact that the minimal surface doesn't reach into the deep IR (the ``boundary" is placed at $p=\e_{\textrm{eff}}$ and any minimal surface in the bulk anchored at the hypersurface $\partial A$ on the boundary actually lives completely on the boundary slice) mirrors the fact that minimal surfaces in the full geometry tend to hug the horizon. This gives the extensive behavior shown in \cite{Swingle:2009wc, Albash:2010mv} by numerical integration in the full geometry.

The electron star is similar but in the near-horizon limit has a Lifshitz form defined by a different scaling for the coefficient $g_{tt}$ \cite{Hartnoll:2011fn}
\be
ds^2=R^2\left(-\frac{1}{r^{2z}}dt^2+\frac{g_{\infty}}{r^2}dr^2+\frac{1}{r^2}dx_i^2\right),
\ee
where $g_{\infty}$ is a constant that depends on various parameters in the construction and $z$ is the dynamical critical exponent. Since we work on constant time slices, the new coefficient $g_{tt}$ is irrelevant, and we see that a constant time slice in the near horizon geometry is simply EAdS$_{d+1}$. This will give the area-law scaling and again no logarithm.  \\

\begin{subsection}
{The OTU Geometry}
\end{subsection}

A geometry with the near-horizon limit AdS$_3\times \mathbb{R}^{d-1}$ is a simple situation in which a logarithmic violation can occur. However, it is possible to have geometries that do not factorize in the IR but produce the correct entanglement entropy scaling for a Fermi surface.\footnote{This analysis was partially carried out during the course of this paper, but a fuller treatment was given in \cite{Ogawa:2011bz}.} We shall present the asymptotically AdS$_4$ geometry in \cite{Ogawa:2011bz} and analyze its singularity nature in Section \ref{singular}.

One begins by picking a metric ansatz
\be
ds^2=\frac{R^2}{z^2}(-f(z)dt^2+g(z)dz^2+dx^2+dy^2)\label{tak}\ee
in four dimensions and requiring the logarithmic behavior with an effective cutoff. According to \cite{Ogawa:2011bz}, this fixes the function $g(z)$ to $z^2/\e_{\textrm{eff}}^2$ deep in the bulk. Since we consider time-independent geometries, $f(z)$ is not fixed by the behavior of the entanglement entropy and is instead partially fixed by imposing the null energy condition in the bulk. We ignore $f(z)$ for now and focus on a given time slice.

As usual, we go to the ``boundary" of the IR geometry (which we identify as $z\rightarrow 0$) and pick a region $A$ defined by the infinite strip: 
\be
A:=x\in(-L_x/2, L_x/2), y\in(-L_y/2, L_y/2)
\ee
with $L_y\gg L_x$. As usual, these are effective lengths $L_{\textrm{eff}}$ seen in the IR when embedded in the full geometry, but we leave off the subscripts to keep the equations less cluttered. The area of the minimal surface is given by 
\be
\textrm{Area}(\gamma_A)=\int dx dy \frac{1}{z}\sqrt{\frac{1}{z^2}+\left(\frac{z'}{\e_{\textrm{eff}}}\right)^2}.
\ee
Translation invariance in $y$ due to the belt geometry allowed us to parametrize the surface as $z(x,y)=z(x)$, with $z'\equiv dz/dx$. We can then perform the $y$-integral to get 
\be
\textrm{Area}(\gamma_A)=L_y\int_{-L_x/2}^{L_x/2}dx\frac{1}{z}\sqrt{\frac{1}{z^2}+\left(\frac{z'}{\e_{\textrm{eff}}}\right)^2}.
\ee
Considering the integrand as a Lagrangian $\mathcal{L}$, then 
\be
H=\frac{\p \mathcal{L}}{\p z'} \cdot z'-\mathcal{L}=-\frac{1}{z^3}\left(\frac{1}{z^2}+\left(\frac{z'}{\e_{\textrm{eff}}}\right)^2\right)^{-1/2}\equiv-\frac{1}{a^2}
\ee
 is a conserved quantity, where we imagine $x$ as a time variable. The minimal surface will be a symmetric function of $x$, which gives $z'(x=0)=0$ for the turning point. We use this conserved quantity to write 
\be
L_x = 2\int_{\e_{\textrm{eff}}}^a \frac{z^3 dz}{\e_{\textrm{eff}}\sqrt{a^4-z^4}} \hspace{10mm} \textrm{Area}=2L_y\int_{\e_{\textrm{eff}}}^a \frac{a^2 dz}{\e_{\textrm{eff}} z\sqrt{a^4-z^4}},
\ee
where we cut off the integrals to extract the divergences. These integrals can be computed analytically and give
\be
L_x=\frac{\sqrt{a^4-\e_{\textrm{eff}}^4}}{\e_{\textrm{eff}}} \hspace{18mm} \textrm{Area}=L_y \frac{\log\left( \frac{a^2+\sqrt{a^4-\e_{\textrm{eff}}^4}}{\e_{\textrm{eff}}^2}\right)}{\e_{\textrm{eff}}}.
\ee
This is one of those rare and delightful cases where expanding for small $\e_{\textrm{eff}}$ is easier without the aid of Mathematica:
\begin{align}
L_x \sim \frac{a^2}{\e_{\textrm{eff}}} \hspace{10mm} \textrm{Area}&\sim L_y\frac{\log\left(\frac{2a^2}{\e_{\textrm{eff}}^2}\right)}{\e_{\textrm{eff}}}\\
\nonumber \\
\implies \textrm{Area}&\sim \frac{L_y}{\e_{\textrm{eff}}}\log \left(\frac{2L_x}{\e_{\textrm{eff}}}\right)\sim \frac{L_y}{\e_{\textrm{eff}}}\log \left(\frac{L_x}{\e_{\textrm{eff}}}\right),
\end{align}
where we have ignored terms subleading in $\e_{\textrm{eff}}$. A similar analysis can be applied to higher dimensional cases. We see that our IR prescription produces the correct scaling in the geometry identified in \cite{Ogawa:2011bz}.\\

\begin{section}
{Singularities of the OTU geometry \label{singular}}
\end{section}

We wish to analyze the singularity nature of the OTU geometry. In the original paper \cite{Ogawa:2011bz}, the authors find an effective gravity embedding of the geometry with fields which become singular at $z=\infty$. We will show that the geometry generically has a null curvature singularity at $z=\infty$ except for a specific set of parameter values. The singularity exists even though all curvature invariants remain finite. In fact, the singularity nature is analogous to that of the Lifshitz geometry, in a sense which we will make precise below. The method of analysis is inspired by \cite{Horowitz:2011gh}, and we will compare our results to those of the Lifshitz geometry, which is included as a special case of our geometry.\\

\begin{subsection}
{Diverging Tidal Forces}\label{div}
\end{subsection}

We begin by considering the metric 

\be
ds^2=\frac{1}{z^2}(-f(z)dt^2+g(z)dz^2+dx^2+dy^2),
\ee
with $f(z)\sim z^{-2m}$ and $g(z)\sim z^{2p}$. The radius of curvature is set to 1. We will restrain from specifying to $p=1$ and $m\geq p$, the conditions required in \cite{Ogawa:2011bz}, until the end. We also note the alternative form of writing the metric which makes the dynamical critical exponent $\gamma$ and hyperscaling violation exponent $\theta$ manifest, for later comparison \cite{Huijse:2011ef, Dong:2012se}:
\be
ds^2=\frac{1}{z^2}\left(-\frac{dt^2}{z^{2d(\gamma-1)/(d-\theta)}}+z^{2\theta/(d-\theta)}dz^2+dx_i^2\right).
\ee
Comparing to our metric shows that $d=2$, $m=2(\gamma-1)/(2-\theta)$ and $p=\theta/(2-\theta)$.

We want to consider the tidal forces felt by an infalling observer, so we consider an orthonormal frame that is parallel propagated along a radial infalling geodesic. The four-velocity is given as $U^{\mu}=(\dot{t},\dot{z},0,0)$, where dots denote derivatives with respect to proper time. We have as an integral of the motion the conserved energy $E=-K_{\mu}U^{\mu}=-g_{tt}\frac{dt}{d\tau}$ and the normalization of the four-velocity gives 

\be
\dot{z}^2=\frac{z^2}{g(z)}\left(-1+\frac{E^2z^2}{f(z)}\right)\label{vel}.
\ee
We write the vielbein parallel propagated along a radially infalling geodesics as 
\begin{align}
(e_0)^{\mu}&=Ez^{2m+2}(\partial_t)^{\mu}+z^{1-p}\sqrt{-1+E^2z^{2+2m}}(\partial_z)^{\mu}\\
(e_1)^{\mu}&=-z^{m+1}\sqrt{-1+E^2 z^{2+2m}}(\partial_t)^{\mu}-Ez^{2+m-p}(\partial_z)^{\mu}\\
(e_i)^{\mu}&=z\left(\partial_{x^i}\right)^{\mu}.
\end{align}
\\
Components of the Riemann tensor in this frame will give the physical tidal forces felt by an infalling observer. We use the constructed vielbein to transform from the ``static" frame to the infalling frame:

\be
R_{abcd}=R_{\mu\nu\rho\sigma}(e_a)^{\mu}(e_b)^{\nu}(e_c)^{\rho}(e_d)^{\sigma},
\ee
and we find the following diverging components in the new frame
\begin{align}
R_{0i0i}&=z^{-2p}(1+p+E^2(m-p)z^{2+2m})\\
R_{1i1i}&=z^{-2p}(-1-m+E^2(m-p)z^{2+2m})\\
R_{0i1i}&=-E(m-p)z^{1+m-2p}\sqrt{(-1+E^2z^{2+2m})}.
\end{align}
Restricting to $m>p\geq0$, the tidal forces diverge as $(m-p)z^{2m-2p+2}$.  Notice that for $p=0$ we simply get the result in \cite{Horowitz:2011gh} which is appropriate for Lifshitz scaling. $m=p=0$ is the nonsingular result of pure AdS. However, in this more general geometry we see that there seem to be other nonsingular results, namely for $m=p$ and $m\leq p-1$. Imposing the null energy condition in the bulk gives the condition $m \geq p$, which leaves only $m=p$ as a nonsingular set of values for the parameters. This constraint translates, in general dimension, to $\gamma=1+\theta/d$ for the condensed matter exponents. In general dimension this also saturates the null energy condition in the bulk. To reproduce the logarithmic violation, we need to fix $p=\theta=1$ in four dimensions and so we are left with the nonsingular metric\footnote{The translation invariance of this metric in the coordinates $x_1$ and $x_2$ suggests that geodesics with momentum in these directions, not considered above, should not produce divergent tidal forces (although this has only been checked explicitly for $R_{0303}$ with momentum in the 2-direction). Furthermore, curvature invariants will all remain finite at $z=\infty$, since in the ``static" frame all components of the Riemann tensor are finite there. The singular cases are similar to the Lifshitz spacetime, in the sense that curvature invariants are all finite while there exists a null curvature singularity. The singularity at $z=0$ is of no importance because this geometry is only valid for $z\gg k_F^{-1}=1$; it is asymptotically AdS, which is well-behaved at $z=0$.}

\be
ds^2=\frac{1}{z^2}\left(\frac{-dt^2}{z^2}+z^2dz^2+dx^2+dy^2\right).
\ee

The hyperscaling violation exponent for this geometry is $\theta=1$ while the dynamical critical exponent is $\gamma=3/2$. Surprisingly, these values occur precisely for gauge theories of non-Fermi liquid states (in a three-loop analysis) \cite{Metlitski:2010pd, Thier:2011gf}. This metric has a Poincare-type horizon at $z=\infty$, to which timelike geodesics can propagate in finite proper time (one can see this by integrating (\ref{vel}); the singularities in the singular geometries are also reached in finite proper time). But the metric should admit an analytic extension of the coordinates, just like the extension from Poincare coordinates to global coordinates in AdS. A quick computation of the surface gravity shows that $z=\infty$ is a degenerate Killing horizon for the Killing vector $\partial_t$. Some intuition for the special status of this metric can be gained by making the coordinate transformation $z=1/r$, after which the metric takes the pleasant form 
\begin{align}
ds^2=-F(r)dt^2+\frac{dr^2}{F(r)}+r^2dx_i^2, \hspace{5mm} F(r)=r^4
\end{align}
The horizon is now located at $r=0$ and we see that a continuation to negative $r$ preserves the signature of the metric.

The singularity for $m>p\geq 0$ is a null curvature singularity, as surfaces of constant $z$ become null as $z\rightarrow \infty$. This can be seen by observing radial null geodesics: $dt=\pm z^{2m+2p}dz$. Since our singularity is at $z=\infty$, it takes an infinite amount of coordinate time to reach the singularity, implying that the singularity is null (recall we work in the regime $m \geq p>0$). 

In higher dimensions, as long as rotational and translational symmetry is kept in all dimensions but $z$ and $t$, the results are similar. The value of $p$ will increase, but the null energy condition will still give $m\geq p$. \\

\begin{subsection}
{Plane Wave Approximation}
\end{subsection}
In this section we show that, as in the Lifshitz geometry, the OTU geometry can be represented near the singularity as a plane wave. To do this analysis we work with the geometry in the form 

\be
ds^2=\left(-r^{2m+2}dt^2+\frac{dr^2}{r^{2p+2}}+r^2(dx_1^2+dx_2^2)\right),
\ee
where we have simply made the redefinition $r=1/z$. Switching to a tortoise coordinate via $dr_*/dr=\sqrt{-g_{rr}/g_{tt}}\implies r_*=-\frac{r^{-m-p-1}}{m+p+1}$ and subsequently lightcone coordinates $u=t-r_*$, $v=t+r_*$ gives the metric in the form 

\begin{align}
ds^2&=\left(\frac{1}{cr_*}\right)^{\frac{2m+2}{c}}(-dt^2+dr_*^2)+\left(\frac{1}{cr_* }\right)^{\frac{2}{c}}(dx_1^2+dx_2^2)\nonumber\\
&=-\left(\frac{4}{c^2(u-v)^2}\right)^{\frac{m+1}{c}}dudv+\left(\frac{4}{c^2(u-v)^2}\right)^{\frac{1}{c}}(dx_1^2+dx_2^2),\nonumber\\
\end{align}
where we have defined $c=m+p+1$. Since $r\sim(r_*)^{-1/c}\sim\left(\frac{1}{u-v}\right)^{1/c}$, we have that small $r$ corresponds to $u\gg v$ if we stick to $c>0$. The viable set of parameters for logarithmic violation of the entanglement entropy is in this range ($m\geq p=1$ in four dimensions). Thus, near the singularity we can approximate the metric as

\be
ds^2\approx -\left(\frac{4}{c^2u^2}\right)^{\frac{m+1}{c}}dudv+\left(\frac{4}{c^2u^2}\right)^{\frac{1}{c}}(dx_1^2+dx_2^2).
\ee
If we make the redefinition

\be
\left(\frac{4}{c^2u^2}\right)^{\frac{m+1}{c}}du=dU,
\ee
then for the coefficient of $dx^2+dy^2$ we get
\be
\left(\frac{4}{c^2}\right)^{1/c}u^{-2/c}=U^{\frac{2}{m+1-p}}\left(\frac{(m+1-p)}{2}\right)^{\frac{2}{m+1-p}},
\ee
which gives a plane wave metric 

\be
ds^2\approx - dUdv+\left(\frac{U(m+1-p)}{2}\right)^{\frac{2}{m+1-p}}(dx_1^2+dx_2^2).
\ee
Notice in the case $p=0, 2m+2=2\gamma$ for critical exponent $\gamma$, we have precisely the Lifshitz metric and the answer we get is precisely the Lifshitz answer, with the coefficient of the second term being given as $(\gamma U/2)^{2/\gamma}$. In general, we can write our oefficient in the suggestive form $(kU/2)^{2/k}$ for $k=m+1-p$; this is precisely the same form as the Lifshitz answer!

To bring the plane wave metric into the standard Brinkmann form, we define
\begin{align}
k=m+1-p, \hspace{9mm}v=V-\frac{X_iX^i}{kU}, \hspace{9mm}x_i=X_i\left(\frac{kU}{2}\right)^{-1/k}
\end{align}
and get 

\be
ds^2\approx -dUdV + dX_i dX^i + W(U)X_i X^i dU^2, \hspace{9mm} W(U)=\frac{1-k}{k^2U^2}.\label{brink}
\ee
This is a good approximation to the geometry near $z=\infty$ as long as $k \neq 1$, since the diverging tidal forces of this geometry and the OTU geometry behave in the same way. If $k=1$, however, the OTU geometry is some nonsingular, asymptotically AdS geometry while the plane wave approximation gives Minkowski space. Thus, the approximation breaks down in this case as there are no diverging tidal forces which dominate the behavior and match up between the two geometries. The metric is appropriate in any number of dimensions, where $p$ will change to give the logarithmic violation. In all cases, however, $p=m\implies k=1$ will be the one nonsingular case which also satisfies the null energy condition.

The calculation of infalling geodesics is identical to that of the OTU geometry. The four-velocity is given as $U^{\mu} = (\dot{U},\dot{V},0,0)$, where dots denote derivatives with respect to proper time. We have as an integral of the motion the energy $E=-K_{\mu}U^{\mu}=\dot{U}/2$. Normalization of the four-velocity gives 
\be
\dot{V}=\frac{1+4E^2W(U)X_iX^i}{2E}.
\ee
We pick the parallel propagated frame
\begin{align}
(e_0)^{\mu}&=2E\left(\partial_U\right)^{\mu}+\frac{1+4E^2W(U)X_iX^i}{2E}\left(\partial_V\right)^{\mu}\\
(e_1)^{\mu}&=2E\left(\partial_U\right)^{\mu}+\frac{-1+4E^2W(U)X_iX^i}{2E}\left(\partial_V\right)^{\mu}\\
(e_i)^{\mu}&=\left(\partial_{X^i}\right)^{\mu}.
 \end{align}
We find diverging components of the Riemann tensor 
\be
R_{0i0i}=R_{1i1i}=R_{0i1i}=\frac{4E^2(k-1)}{k^2U^2}=(m-p)z^{2(m+1-p)},
\ee
which diverges precisely like the original metric as $z\rightarrow \infty$.\\

\begin{subsection}
{A Stringularity}
\end{subsection}

Studying test strings in the plane wave background (\ref{brink}) is a solved problem; the strings propagating through the singularity become infinitely excited, suggesting that the geometry is singular in the string theoretic sense, i.e. ``stringular." However, without a string theory embedding this should be viewed merely as a hint; as is well known by now there are other effects, e.g. the dynamics of D-branes, which can resolve singularities. 

The mathematics of these test strings is identical to \cite{Horowitz:2011gh}, which in turn is essentially identical to \cite{deVega:1990ke}. We will calculate the expectation value of the mass squared and mode number operators to see that they diverge as long as $m \neq p$. This more general case for the value of $p$ is relevant for higher dimensions, as imposing the null energy condition in the bulk always gives $m\geq p$.

We begin with the usual Polyakov action describing strings propagating in a general curved background
\be
S=-\frac{1}{4\pi\alpha'}\int d\tau d\sigma \sqrt{-\gamma}\gamma^{ab}g_{\mu\nu}(X)\partial_a X^{\mu}\partial_b X^{\nu}.
\ee
We can fix some of the gauge redundancy by working in conformal gauge $\gamma_{ab}=e^{\phi(\sigma,\tau)}\eta_{ab}$. The residual freedom (changing $\sigma$ and $\tau$ by solutions to the wave equation) is fixed by imposing light cone gauge, $U=\alpha' p \tau$. Fourier expanding the embedding coordinates as 
\be
X^i (\sigma,\tau)=\sum_{n=-\infty}^{n=+\infty}X^i_n(\tau)e^{in\sigma}
\ee
allows us to write the equations of motion in this gauge as 
\begin{align}
&\ddot{X}_n^i+\left[n^2-\alpha'^2p^2W(\alpha'p\tau)\right]X_n^i=0 \label{EOM}\nonumber\\
&\alpha' p \dot{V}=(\dot{X}^i)^2+(X'^i)^2+\alpha '^2p^2W(\alpha'p\tau)X_iX^i\\
&\alpha'pV'=2\dot{X}^iX'^i,\nonumber
\end{align}
where we see that the equation of motion for the embedding coordinates is a Schrodinger equation.\footnote{This Schrodinger equation has potential $\alpha'^2 p^2 W(\alpha' p \tau)=\frac{1-k}{U^2k^2}$ near the singularity. Notice that, contrary to the statements in \cite{deVega:1990ke} which ignore the strength of the potential, for the cases of interest $k\geq 1$ and the strength of this inverse square potential is bounded between $-1/4$ and $0$, allowing for transitions in the quantum mechanics problem; thus, the string can make it through the singularity at $\tau=0$ and our matching is justified.}   From this equation we see that the mode functions depend on the combination $n\tau$:
\be
X_n^i(\tau)=X_1^i(n\tau)\label{indep}.
\ee
We consider a plane wave with flat spacetime before and after it. In other words, we take 
\begin{align}
W(U)=\begin{cases}0  \hspace{5mm} &\mathrm{if }\hspace{1mm}|U|\geq T\\
\frac{1-k}{k^2} \left(\frac{1}{U^2}-\frac{1}{T^2}\right) \hspace{5mm} &\mathrm{if }\hspace{1mm} |U|\leq T \end{cases}
\end{align}
for some large $T$, where $\frac{1-k}{k^2}$ is negative for the singular cases of interest ($m>p$), giving an attractive potential in our Schrodinger equation. For the regions $|U|\geq T$ we can do the standard flat space expansions of the embedding coordinates:
\begin{align}
X^i(\sigma,\tau)&=q^i_<+2\alpha'p^i_<\tau+\sum_{n\neq 0}e^{in\sigma}X_{n<}^i(\tau)  &&\tau \leq -\tau_0\label{fse<}\\
X^i(\sigma,\tau)&=q^i_>+2\alpha'p^i_>\tau+\sum_{n\neq 0}e^{in\sigma}X_{n>}^i(\tau)  &&\tau \geq -\tau_0\label{fse>}\\
X_{n>,<}^i(\tau)&=i\frac{\sqrt{\alpha'}}{n}(a_{n>,<}^ie^{-in\tau}-\tilde{a}_{-n>,<}^ie^{in\tau}) &&n \neq 0
\end{align}
for $\tau_0 = T/\alpha' p$ and mode operators $a_{n>,<}^i$, $\tilde{a}_{n>,<}^i$ which satisfy the canonical commutation relations. The slightly unconventional normalization has been chosen to conform to the past literature on the subject. The Bogoliubov transformations connecting these two expansions are written generally as 
\begin{align}
a_{n>}^i &= A_n^ia_{n<}^i+B_n^i\tilde{a}_{n<}^{i\dagger}\label{BOG}\\
\tilde{a}_{n>}^i &= A_n^i\tilde{a}_{n<}^i+B_n^i\tilde{a}_{n<}^{i\dagger}.\nonumber
\end{align}

The Schrodinger equation for $X_{n<}^i$ can be solved exactly by Bessel functions and expanded as $\tau\rightarrow 0^-$ to get 
\be
X_{n<}^i(\tau)=D_n^i|n\tau|^{(1-\nu)/2}+ F_n^i|n\tau|^{(1+\nu)/2} \hspace{5mm}\textrm{for}\hspace{2mm} \nu=\sqrt{1-4C(1-k)/k}.\ee

The problem is essentially a scattering problem off of the given potential. Accordingly, we pick as an initial state
\be
f_n^i(\tau)=e^{in\tau}\hspace{5mm}\textrm{for} \hspace{2mm}\tau<-\tau_0
\ee
and get a solution of the equation of motion for the embedding coordinate as
\be
f_n^i(\tau)=e^{in\tau}+\frac{p^2\alpha'^2}{2in}\left(e^{in\tau}\int_{-\infty}^{\tau}d\tau'e^{-in\tau'}f_n^i(\tau')W(\alpha'p\tau')-e^{-in\tau}\int_{-\infty}^{\tau}d\tau'e^{in\tau'}f_n^i(\tau')W(\alpha'p\tau')\right).
\ee
One can verify this is a solution by plugging into (\ref{EOM}). From the equations of motion for our embedding coordinates (\ref{fse<}), (\ref{fse>}) and the Bogoliubov transformation (\ref{BOG}) we get 
\be
B_n^i=\frac{p^2\alpha'^2}{2in}\int_{-\tau_0}^{\tau_0}d\tau e^{in\tau}f_n^i(\tau)W(\alpha'p\tau).\label{BOGsol}
\ee

The mass and number operators in the $U\geq T$ region are given by 
\begin{align}
M^2_{>}&=\frac{1}{\alpha'}\sum_{n=1}^{\infty}\left(a_{n>}^{i\dagger}a_{n>}^i+\tilde{a}_{n>}^{i\dagger}\tilde{a}_{n>}^i\right) +m_0^2,\\
N_>&=\sum_{n=1}^{\infty}\frac{1}{n}\left(a_{n>}^{i\dagger}a_{n>}^i+\tilde{a}_{n>}^{i\dagger}\tilde{a}_{n>}^i\right),
\end{align}
where $m_0^2=-(D-2)/(12\alpha')$ is the tachyonic mass in this normalization.

Using the Bogoliubov transformations (\ref{BOG}), we can find the expectation values of the mass and number operators in the ingoing ground state as
\begin{align}
\langle M^2_> \rangle = \frac{\langle 0_<|M_>^2|0_<\rangle }{\langle 0_<|0_<\rangle}=m_0^2+\frac{2}{\alpha'}\sum_{n=1}^{\infty}\sum_{i=1}^{D-2} n|B_n^i|^2,\\
\langle N_>\rangle=\frac{\langle 0_<|N_>|0_<\rangle }{\langle 0_<|0_<\rangle}=2\sum_{n=1}^{\infty}\sum_{i=1}^{D-2}n|B_n^i|^2,
\end{align}
for $D$ spacetime dimensions. Making the change of variables $y=n\tau$ in (\ref{BOGsol}) and using (\ref{indep}) and the fact that $T$ is large gives
\[B_n^i = \frac{C(1-k)}{2ik^2}\int_{-n\tau_0}^{n\tau_0}\frac{e^{iy}f_1^i(y)}{y^2}\approx \lim_{n\rightarrow\infty} i\frac{C(k-1)}{2k^2}\int_{-n\tau_0}^{n\tau_0} dy \frac{e^{iy}f_1^i(y)}{y^2}\]
This integral is finite and independent of $n$ for $k\neq 1$ \cite{Horowitz:2011gh}. Thus, each mode is excited equally and mass and number operators diverge, leaving the door open for a full-fledged stringularity. We are only saved when $k=1$, which corresponds to $m=p$. This is precisely the case where there is no general relativistic singularity. We stress again that this calculation is intended to be suggestive rather than a definitive remark about the behavior of this geometry in string theory.\\

\begin{section}
{Effective Gravity Embedding of Nonsingular Geometry \label{embedding}}
\end{section}
The OTU geometry for the singular $m>p=1$ case is derived from Einstein-Maxwell-dilaton theory in \cite{Ogawa:2011bz}. We now demonstrate a solution in the full geometry for the nonsingular case $m=p=1$. The geometry we want to produce is $ds^2=\frac{1}{z^2}\left(-f(z)dt^2+g(z)dz^2+dx^2+dy^2\right)$ with $f(z)=1/(1+z^2/k)$ and $g(z)=1+z^2/z_F^2$. This has the required nonsingular IR behavior which produces the logarithmic violation of entanglement entropy, in addition to being asymptotically $AdS_4$. It satisfies the null energy condition everywhere in the bulk when $k<z_F^2$, which is necessary for the reality of the scalar field solution given by (\ref{phi}).

\begin{figure}[htb]
\centering
\includegraphics[width=0.78\textwidth]{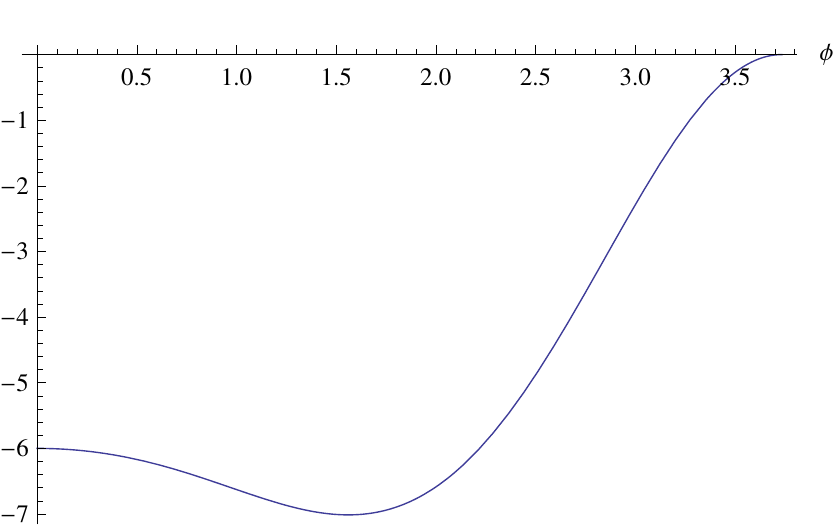}
\caption{$V(\phi)-6=\frac{\mathrm{sc}[\phi/(2 \sqrt{c_1}); 
    c_1]^2 (c_2 +c_3 \mathrm{sc}[\phi/(2 \sqrt{c_1}); 
      c_1]^2 + c_4 \mathrm{sc}[\phi/(2 \sqrt{c_1}); c_1]^4 + 
     c_5 \mathrm{sc}[\phi/(2 \sqrt{c_1}); c_1]^6)}{(1 + 
     \mathrm{sc}[\phi/(2 \sqrt{c_1}); c_1]^2)^2 (z_F^2 + 
     k \mathrm{sc}[\phi/(2 \sqrt{c_1}); c_1]^2)^2}-6$, with $k=1$ and $z_F=2$. The potential vanishes in the IR.}
\label{fig:Vofphi}
\end{figure}

\begin{figure}[htb]
\centering
\includegraphics[width=0.80\textwidth]{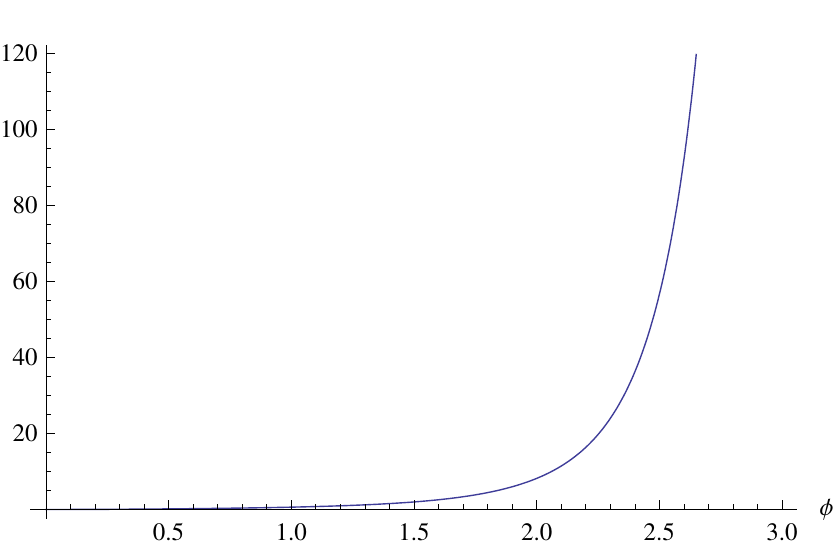}
\caption{$Z(\phi)=\frac{2 k^2 \mathrm{sc}[\phi/(2 \sqrt{c}); 
    c]^2 (1 + \mathrm{sc}[\phi/(2 \sqrt{c}); c]^2)^2 (z_F^2 + 
     k \mathrm{sc}[\phi/(2 \sqrt{c}); c]^2)^2}{z_F^2 (z_F^2 + 
     2 (k + 2 z_F^2) \mathrm{sc}[\phi/(2 \sqrt{c}); c]^2 + 
     5 k \mathrm{ sc}[\phi/(2 \sqrt{c}); c]^4)}$, with $k=1$ and $z_F=2$. The divergence is simply indicating that the gauge coupling, which scales as $Z(\phi)^{-1}$, is vanishing in the IR.}
\label{fig:Wofphi}
\end{figure}

We begin with the action 
\be
S=\frac{1}{2\kappa}\int\sqrt{-g}\left[(R-2\Lambda)-Z(\phi)F^{\mu\nu}F_{\mu\nu}-\frac{1}{2}\partial_{\mu}\phi\partial^{\mu}\phi-V(\phi)\right].
\ee
With an ansatz $\phi=\phi(z)$ and $A_t=a(z)$, we get $a'(z)=\frac{A}{Z(\phi)}\sqrt{f(z)g(z)}$ for an integration constant $A$. All equations of motion can be solved by

\begin{align}
V(\phi)\nonumber\\=&\frac{z^2g(z)f'(z)^2-4f(z)^2(6g(z)-6g(z)^2+zg'(z))+zf(z)(zf'(z)g'(z)+g(z)(8f'(z)-2zf''(z)))}{4f(z)^2g(z)^2},\nonumber\\
\nonumber\\
Z(\phi)&=\frac{-8A^2z^3f(z)^2g(z)^2}{zg(z)f'(z)^2+f(z)(zf'(z)g'(z)+g(z)(4f'(z)-2zf''(z)))},\nonumber\\
\nonumber\\
\phi'(z)&=\sqrt{\frac{-2(g(z)f'(z)+f(z)g'(z))}{zf(z)g(z)}}.
\end{align}
We place this preliminary form to highlight the ease with which metrics can be produced with this action and ansatz. For the specific $f(z)$ and $g(z)$ proposed above, we find 
\begin{align}
A_t(z)&=\frac{Az_F\sqrt{k}}{2z(k+z^2)^{3/2}\sqrt{z^2+z_F^2}}\\
\phi(z)&=2\sqrt{1-k/z_F^2}F\left[\arctan(z/\sqrt{k})\Big |1-k/z_F^2\right]\label{phi},
\end{align}
where $F[a|m]$ is the elliptic integral of the first kind. Constants $c_i$ have been defined to condense the expression; they are just combinations of $z_F$ and $k$. The dilaton is plotted in Figure \ref{fig:phi}. We can invert this to get $z$ in terms of $\phi$ and get expressions for $V(\phi)$ and $Z(\phi)$:

\begin{align}
V(\phi)&=\frac{\mathrm{sc}[\phi/(2 \sqrt{c_1}); 
    c_1]^2 (c_2 +c_3 \mathrm{sc}[\phi/(2 \sqrt{c_1}); 
      c_1]^2 + c_4 \mathrm{sc}[\phi/(2 \sqrt{c_1}); c_1]^4 + 
     c_5 \mathrm{sc}[\phi/(2 \sqrt{c_1}); c_1]^6)}{(1 + 
     \mathrm{sc}[\phi/(2 \sqrt{c_1}); c_1]^2)^2 (z_F^2 + 
     k \mathrm{sc}[\phi/(2 \sqrt{c_1}); c_1]^2)^2},\nonumber\\
\nonumber\\
Z(\phi)&=\frac{2 k^2 \mathrm{sc}[\phi/(2 \sqrt{c}); 
    c]^2 (1 + \mathrm{sc}[\phi/(2 \sqrt{c}); c]^2)^2 (z_F^2 + 
     k \mathrm{sc}[\phi/(2 \sqrt{c}); c]^2)^2}{z_F^2 (z_F^2 + 
     2 (k + 2 z_F^2) \mathrm{sc}[\phi/(2 \sqrt{c}); c]^2 + 
     5 k \mathrm{ sc}[\phi/(2 \sqrt{c}); c]^4)},
\end{align}
where sc$[a;b]$ is the Jacobi elliptic function, which appears due to inverting the elliptic integral in (\ref{phi}). These are not illuminating forms of the functions; their more tangible plots can be found in Figures \ref{fig:Vofphi} and \ref{fig:Wofphi}. We see that the dilaton asymptotes to a constant, whereas in the singular cases where $m>1$ it diverges logarithmically. As one can guess, this constancy of the dilaton in the IR is robust to modifications of the metric which keep the asymptotic behaviors the same.
\\
\\
\begin{figure}[htb]
\centering
\includegraphics[width=0.75\textwidth]{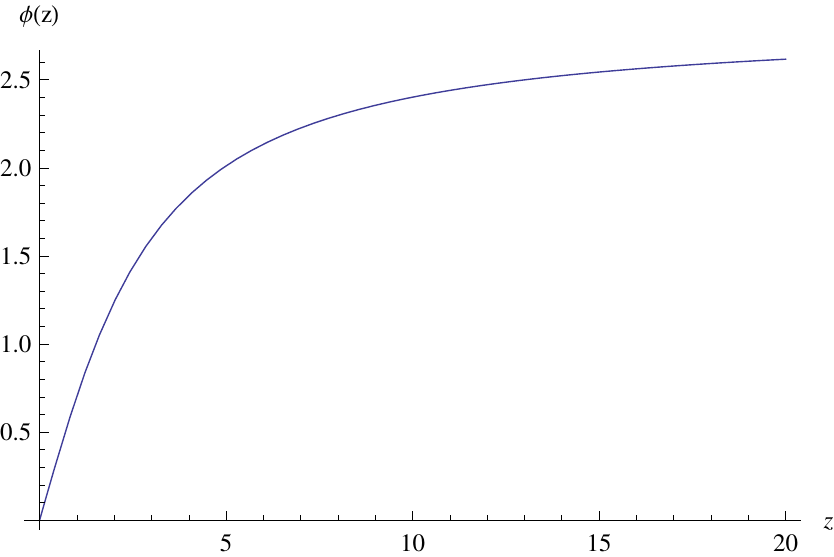}
\caption{$\phi(z)=2\sqrt{1-k/z_F^2}F\left[\arctan(z/\sqrt{k})\Big |1-k/z_F^2\right]$, with $k=1$ and $z_F=2$. The scalar field now asymptotes to a constant in the IR, whereas in the case $m>p=1$ it diverges logarithmically.}
\label{fig:phi}
\end{figure}

\begin{figure}[htb]
\centering
\includegraphics[width=0.78\textwidth]{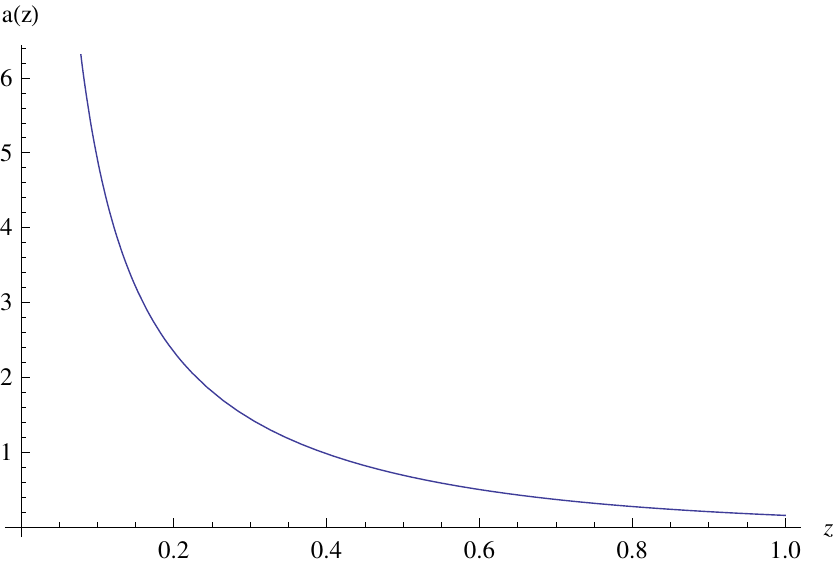}
\caption{$a(z)=\frac{Az_F\sqrt{k}}{2z(k+z^2)^{3/2}\sqrt{z^2+z_F^2}}$}
\label{fig:phi}
\end{figure}

\begin{section}
{Conclusions}\label{conclusions}
\end{section}
In this paper, we analyzed situations in which the logarithmic violation of the entanglement entropy, indicative of a theory with a Fermi surface, is expected to occur holographically. This analysis uses weakly coupled gravity in the bulk, which means we are assuming that the logarithmic violation persists to strong coupling. The AdS$_3 \times \mathbb{R}^{d-2}$ geometry was presented as a simple case which produces the violation, though the dual interpretation is unclear.

The geometry of \cite{Ogawa:2011bz} was presented as a more general form which can produce the logarithmic violation. It has gained support from analyzing the scaling of the thermal entropy density and it has been argued that the holographic entanglement entropy formula only detects ``hidden" Fermi surfaces \cite{Huijse:2011ef}. Further holographic explorations are performed in \cite{Dong:2012se}. This class of geometries generically has a null curvature singularity at $z=\infty$ which is reached in finite time, except for when $m=p$. In this case, the surface $z=\infty$ is like the Poincare horizon of $AdS$, and the metric can be analytically extended beyond. In four dimensions, this metric takes the form
\be
ds^2=\frac{1}{z^2}\left(\frac{-dt^2}{z^2}+z^2dz^2+dx_1^2+dx_2^2\right).
\ee
We can gain further intuition for this candidate by writing down the general metric with dynamical critical exponent $\gamma$ and hyperscaling violation exponent $\theta$ \cite{Huijse:2011ef, Dong:2012se}
\be
ds_{d+2}^2=\frac{1}{z^2}\left(-\frac{dt^2}{z^{2d(\gamma-1)/(d-\theta)}}+z^{2\theta/(d-\theta)}dz^2+dx_i^2\right)
\ee
The logarithmic violation of entanglement entropy occurs for $\theta=d-1$ which equals 1 in our case of 4 bulk dimensions. We then see that $\gamma=3/2$, which saturates the bound from the null energy condition. It is interesting to note that
these specific values of the exponents are what arise in a three-loop analysis of gauge theories of non-Fermi liquid states \cite{Metlitski:2010pd, Thier:2011gf}. More generally, the nonsingular cases are given by $\gamma=1+\theta/d$, which again saturates the null energy condition.

The singular cases are only an obstruction to studying the far IR of the theories, as there exist black hole solutions which mask the singularities; in this sense, they are ``good" by the classification of Gubser \cite{Gubser:2000nd} and the limit of $T\rightarrow 0$ can be taken, which corresponds to the finite temperature IR cutoff being removed in the dual theory. However, to study the $T=0$ point, the solution with $m=p$ should be employed to avoid a naked singularity. An embedding of this geometry into Einstein-Maxwell-dilaton theory has been provided in Section \ref{embedding}. The action considered does not permit a consistent IR truncation (because the dilaton asymptotes to a constant), but if a different action with such a consistent IR truncation is found, then black hole solutions with specific heat $C\sim T^{2/3}$ are guaranteed to exist \cite{Iizuka:2011hg, Swingle:2011mk, Dong:2012se, Huijse:2011ef, Ogawa:2011bz}, allowing you to study the system consistently both at zero and finite temperature.\\

\acknowledgments
I would like to thank Dionysios Anninos, Xi Dong, Sean Hartnoll, Nabil Iqbal, Per Kraus, Subir Sachdev, Josh Samani, Steve Shenker, Gonzalo Torroba, and especially Shamik Banerjee for helpful discussions. I would also like to thank my advisor Leonard Susskind for guidance and resources. I am supported by the Stanford Institute for Theoretical Physics under NSF Grant PHY-0756174.
\appendix
\section{Holographic Entanglement Entropy Computations}\label{holo}

In this appendix we apply the proposal to compute the entanglement entropy in a CFT holographically. We show how the area-law scaling comes naturally from the asymptotically AdS geometry. We begin with the logarithmic scaling in three bulk dimensions and move to higher dimensions. We end with a consideration of the extensive scaling in flat space.\\

\subsection{Entanglement Entropy in 2D CFT}

We will work in Poincare coordinates 
\be
ds^2=\frac{R^2}{z^2}(-dt^2+dz^2+dx^2).
\ee
The boundary theory is 1+1 dimensional, so the region $A$ is an interval of length $L$ at a fixed time slice $t=t_0$. It is divided from region $B$ by the points $x=-L/2$ and $x=L/2$. The minimal surface in this dimensionality reduces to a spacelike geodesic. To get finite answers we put the boundary theory at the slice $z=\e$. The geodesic $\gamma_A$ can be parametrized by the non-affine parameter $s$ as 
\be
(z,x)=\frac{L}{2}(\sin s, \cos s).
\ee
Computing the length of this geodesic, we get 
\be
\textrm{Length}(\gamma_A)=\int_{\epsilon_0}^{\pi-\epsilon_0}\sqrt{g_{ij}\frac{dx^{i}}{ds}\frac{dx^{j}}{ds}}ds=\int_{\epsilon_0}^{\pi-\epsilon_0}\frac{Rds}{\sin s}=2R\log \frac{L}{\e}.
\ee
where $\epsilon_0=2\e/L$ and $g_{ij}$ is the induced metric on a constant time slice. This gives for the entanglement entropy 
\be
S_A = \frac{\textrm{Length}(\gamma_A)}{4G_N}=\frac{R}{2G_N}\log \frac{L}{\e}.
\ee
This reproduces the logarithmic scaling we expected.

We see in this computation the famous UV-IR connection; an IR cutoff in the bulk (picking a slice a finite but large distance away, $z=\e$) corresponds to a UV cutoff in the boundary theory.\\

\subsection{Entanglement Entropy in Higher Dimensions}
 
We work again in the Poincare coordinates 

\be
ds^2=\frac{R^2}{z^2}(-dt^2+dz^2+dx_i^2)\ee
for $i=1,2,\dots, d-1$. In higher dimensions there are more shapes we can pick for the region $A$ on the boundary $z=\e$. The computation of the minimal surface simplifies when a region is picked that maintains as much of the symmetry as possible. We will pick the simplest such surface, the ``belt" geometry, which has $x_1\in (-L/2,L/2), x_i \in (-L_{\perp}/2,L_{\perp}/2)$ for $i=2,3,\dots, d-1$ with $L_{\perp}\gg L$. In this case, we minimize the area functional 
\be
R^dL_{\perp}^{d-1}\int_{-L/2}^{L/2}\frac{\sqrt{1+(dz/ds)^2}}{z^d}ds
\ee
and get 
\be
S_A=\frac{R^d}{2G_N(d-1)}\left(\frac{L_{\perp}}{\e}\right)^{d-1}+\textrm{subleading terms,}
\ee
which is the expected area law behavior.
\\
 
\subsection{Comparison to Flat Space Scaling}\label{flat}
 
Here we consider the scaling of entanglement entropy in flat space. The minimal surfaces are simple to understand, and one can quickly see that for a given slicing of flat space, one gets extensive scaling of the minimal surface, i.e. the minimal surface in the bulk anchored at $\partial A$ on the boundary scales as the volume of $\partial A$ as opposed to its area. This can be simply understood to be coming from the fact that a minimal surface connecting a large $d$-dimensional circle (the shape is not relevant; we pick a circle for concreteness) in flat space is a $d+1$-dimensional ball, which scales extensively. We contrast this extensive scaling $S_A \sim L^{d}/\e^d$ in $\mathbb{R}^{d+1,1}$ with the AdS$_{d+2}$ result $S_A \sim L^{d-1}/\e^{d-1}$. This distinction will be important below.\\

\section{Working in the Full Geometry}\label{fullgeometry}
A satisfying way to investigate whether a logarithm appears is to work in the full geometry and pick large surfaces that probe deep into the bulk. In this case, the area of this large surface will come with the usual UV divergent area-law piece (which accompanies any asymptotically AdS spacetime) but will have additional pieces that correspond to the surface reaching deep into the bulk. It is possible to identify an effective cutoff and effective length that appear, $\e_{\textrm{eff}}$ and $L_{\textrm{eff}}$, and look for a piece
\be
\frac{L_{\textrm{eff}}^{d-1}}{\e_{\textrm{eff}}^{d-1}}\log\frac{L_{\textrm{eff}}}{\e_{\textrm{eff}}}.
\ee
We illustrate this behavior with an illuminating toy example constructed in \cite{Albash:2011nq}; there, the authors consider a bulk geometry consisting of two AdS pieces patched together with a sharp domain wall
\be
ds^2=e^{2A(r)}(-dt^2+dx_i^2)+dr^2,
\ee
where
\be
A(r)=\begin{cases}
r/R_{UV} & \textrm{for $r>r_{DW}$}\\
r/R_{IR} & \textrm{for $r<r_{DW}$}
\end{cases}
\ee
This is a discontinuous version of realistic models with a continuous domain wall, like the RG flow from $\mathcal{N}=4$ Super-Yang Mills to an $\mathcal{N}=1$ fixed point (called the Leigh-Strassler fixed point \cite{Leigh:1995ep}) triggered by a mass term for one of the chiral multiplets. Continuing with the toy model in AdS$_5$, we find a minimal surface anchored on the boundary on a ball of radius $l$. The area of the minimal surface is found to be 
\begin{align}
\frac{\textrm{Area}}{4\pi}=&\frac{R_{UV}^3}{2}\left(\frac{l^2}{\e^2}+\log\left(\frac{\e}{l}\right)\right)+\frac{R_{UV}^3}{4}(1-2\log(2))\nonumber\\
&-\frac{1}{2}e^{2r_{DW}/{R_{UV}}}R_{UV}l\sqrt{l^2-\tilde{l}_{cr}^2}+\frac{1}{2}R_{UV}^3\tanh^{-1}\left(\frac{\sqrt{l^2-\tilde{l}_{cr}^2}}{l}\right)\nonumber\\
&+\frac{R_{IR}}{2}e^{2r_{DW}/R_{IR}}\sqrt{l^2-\tilde{l}_{cr}^2}\sqrt{l^2-\tilde{l}_{cr}^2+R_{IR}^2 e^{-2r_{DW}/R_{IR}}}\nonumber\\
&-\frac{R_{IR}^3}{2}\tanh^{-1}\left(\sqrt{\frac{l^2-\tilde{l}_{cr}^2}{l^2-\tilde{l}_{cr}^2+R_{IR}^2e^{-2r_{DW}/R_{IR}}}}\right)+\mathcal{O}(\e),\label{long}
\end{align}
where the definition 
\be
\tilde{l}_{cr}^2=R_{UV}^2e^{-2r_{DW}/R_{UV}}=l_{cr}^2+\mathcal{O}(\e^2)
\ee
has been made. The first line in (\ref{long}) would be the answer if we were in pure AdS. There are also purely IR terms and IR-UV mixing terms. Identifying 
\be
L_{\textrm{eff}}^2=l^2-l_{cr}^2+\mathcal{O}(\e^2), \hspace{12mm} \e_{\textrm{eff}}=R_{IR}e^{-2r_{DW}/R_{IR}}
\ee
and expanding the terms in (\ref{long}) proportional to $R_{IR}^3$ for $\e_{\textrm{eff}}/L_{\textrm{eff}}\ll 1$, the authors get an IR contribution 
\be
\frac{R_{IR}^2}{2}\left(\frac{L_{\textrm{eff}}^2}{\e_{\textrm{eff}}^2}+\log\left(\frac{\e_{\textrm{eff}}}{L_{\textrm{eff}}}\right)\right)+\frac{R_{IR}^3}{4}(1-2\log(2))+\mathcal{O}(\e_{\textrm{eff}}/L_{\textrm{eff}}),
\ee
which is the IR version of the pure AdS answer. We thus see how an effective length and cutoff appear in terms of sensible parameters of the theory and give the expected answer in the IR. In the realistic models, the minimal surface continues to break up in a sensible way, with the piece on one side of the domain wall giving information relevant to IR physics and the piece on the other side of the domain wall giving information relevant to UV physics.

This toy example is an illustration that the minimal surface of the full geometry is smarter than it looks. When we excise the UV geometry and work solely in the IR geometry in our analysis, we are imagining our answers as being the IR pieces coming from a more complete procedure like the one above. Of course, important UV-IR mixing terms are not picked up by this analysis.
\\

\bibliographystyle{JHEP}
\bibliography{fermibibliography}

\end{document}